\documentclass[journal]{IEEEtran}
\usepackage{graphicx} 
\usepackage{algorithm}
\usepackage{algorithmic}
\usepackage{comment}
\usepackage{amsmath}
\usepackage{multirow}
\usepackage{caption}
\usepackage{color}

\ifCLASSINFOpdf
\else
\fi
\begin{document}
%
\title{Multilayer Graph Contrastive Clustering Network}
%
%
%

\author{Liang~Liu,
        Zhao~Kang,
        Ling~Tian,
        Wenbo~Xu,
        and~Xixu~He
\thanks{L. Liu, Z. Kang and L. Tian are with the Department
of Computer Science and Engineering, University of Electronic
Science and Technology of China, Chengdu 611731, China (e-mail: liuliang.uestc@gmail.com; zkang@uestc.edu.cn; lingtian@uestc.edu.cn).}
\thanks{W.B. Xu and X.X He are with the Department of Resources and Environment, University of Electronic Science and Technology of China, Chengdu 611731, China (e-mail: xuwenbo@uestc.edu.cn; HL@uestc.edu.cn).}
}

\maketitle

\begin{abstract}
Multilayer graph has garnered plenty of research attention in many areas due to their high utility in modeling interdependent systems. However, clustering of multilayer graph, which aims at dividing the graph nodes into categories or communities, is still at a nascent stage. Existing methods are often limited to exploiting the multiview attributes or multiple networks and ignoring more complex and richer network frameworks. To this end, we propose a generic and effective autoencoder framework for multilayer graph clustering named Multilayer Graph Contrastive Clustering Network (MGCCN). MGCCN consists of three modules: (1)Attention mechanism is applied to better capture the relevance between nodes and neighbors for better node embeddings. (2)To better explore the consistent information in different networks, a contrastive fusion strategy is introduced. (3)MGCCN employs a self-supervised component that iteratively strengthens the node embedding and clustering. Extensive experiments on different types of real-world graph data indicate that our proposed method outperforms state-of-the-art techniques.
\end{abstract}

\begin{IEEEkeywords}
Multilayer graph, multiple networks, contrastive clustering, self-supervised.
\end{IEEEkeywords}

%
\IEEEpeerreviewmaketitle

\section{Introduction}
%
%
%
%






Graph or network has been a popular data structure to represent the interdependency among objects, such as drug-drug interactions, the friendship relation among people, the citation relation among papers. Graph clustering, which unveils the meaningful patterns of node grouping, is a fundamental task for graph mining in the areas of sociology, biology, and computer science. Emerging applications in security and resilience of critical infrastructures to natural hazards, terrorist activities, and cyberthreats involve multiple types of relationships sharing the same set of nodes, which results in a recent surge of interest in multilayer graph analysis \cite{yadav2020resilience,bermudez2019twitter,baroud2015inherent}. Their study objective is to find which segments of the multilayer graph are vulnerable to a particular hazard and isolate unhealthy components \cite{yuvaraj2021topological}. 
An academic network, another example of multilayer graph, can denote the papers by nodes and capture different types of relationships (paper with paper relationship, paper and author relationship) with multiple layers of connectivity. 

Clustering \cite{leng2019graph} of multilayer graph partitions nodes into communities by accounting for the different relation types. Contrary to unilayer graph, it is still at a nascent stage \cite{interdonato2020multilayer}. Most existing approaches adapt conventional unilayer methods to the multilayer scenario, e.g., stochastic block models \cite{wilson2017community}, layer aggregation techniques \cite{mercado2019spectral,paul2020spectral,chen2017multilayer,GhecheMultilayerNetwork2020}. These methods are limited due to their shallow architecture. Nowadays, many graphs are with node attributes/features. Furthermore, the attributes 
could be different for nodes in different layers. It has been shown that attributes themselves also contain discriminative information, which helps to discover the potential relations among nodes \cite{lin2021graph}. For example, two papers containing textual features (e.g., abstracts, keywords) possibly belong to the same area in the academic graph. The complex non-Euclidean graph structures and various features pose a number of new challenges for multilayer graph clustering. First, how to jointly capture the topology structure and feature information. Second, how to effectively 
aggregate the heterogeneous multiview/multilayer information.
\begin{figure*}[!htbp]
    \centering
    \includegraphics[width=0.9\textwidth]{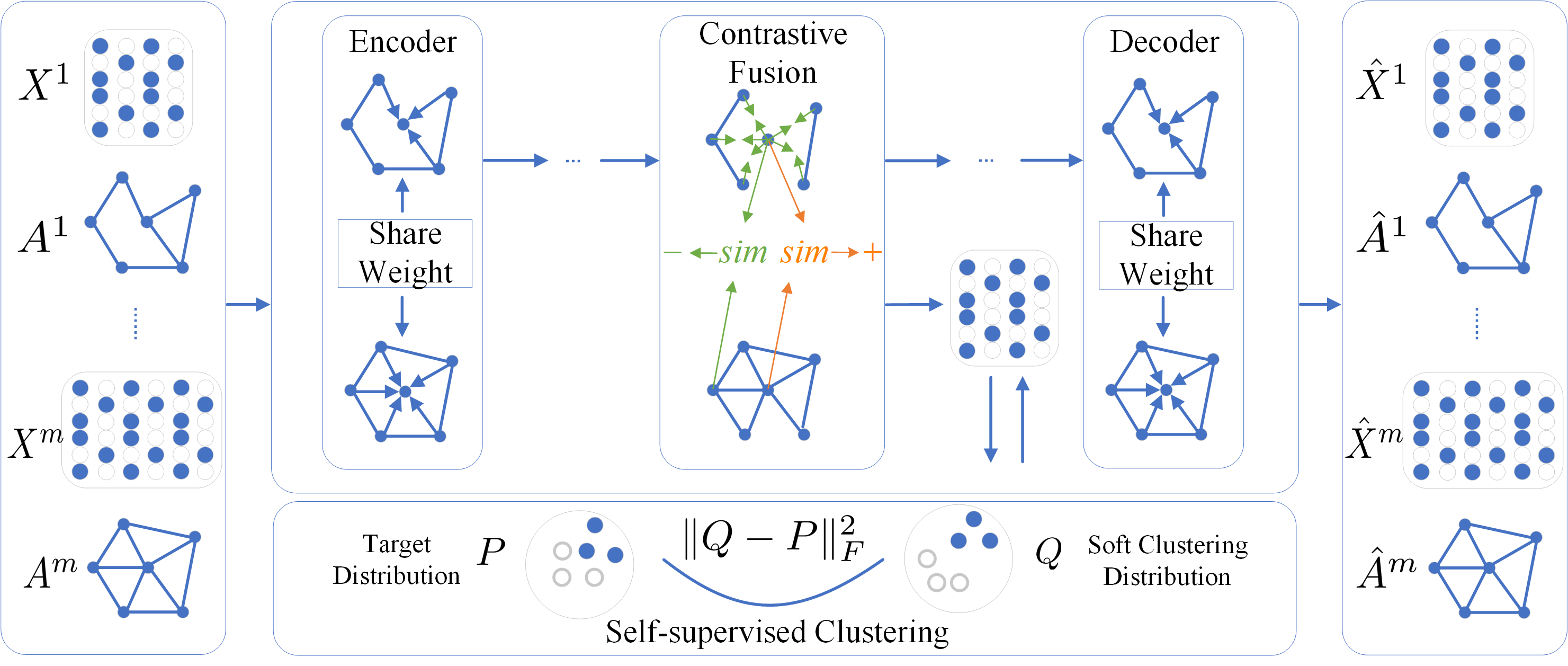}
    \caption{MGCCN framework. The model consists three modules: 1) Self-Reconstruction learns the embedding of each layer by capturing structure and content information. 2) Contrastive fusion captures the consistent information in different layers by pulling close positive pairs and pushing away negative pairs in intra-layer and inter-layer. 3) Self-supervised clustering iteratively boosts the quality of embeddings and clustering.}
    \label{fig:model}
\end{figure*}

Recently, Graph Convolutional Networks (GCNs) have made great progress in various graph learning tasks \cite{liu2021sampling}, such as graph classification, link prediction, and node clustering. These approaches are mostly inspired by graph auto-encoder (GAE) and variatioal GAE (VGAE) \cite{kipf2016variational}, which comprise a GCN encoder and a reconstruction decoder to embed the graph and attributes.
To cluster multilayer graph, Fan et al. \cite{fan2020one2multi} propose to reconstruct multiple graphs based on the embedding of an informative graph chose by modularity. It assumes that all layers share the common attributes and can not make full use of multiview attributes. On the other hand, Cheng et al. \cite{cheng2020multi} design a clustering network for multiview attribute graph, where all views share a common graph. Therefore, clustering of multilayer graph while accounting for multiview attributes remains unexplored. 

To address these challenges, we introduce a generic and effective GCN framework named Multilayer Graph Contrastive Clustering Network (MGCCN). The overall pipeline of MGCCN is shown in Fig \ref{fig:model}. To achieve a consistent cluster pattern in different layers, MGCCN consists of three modules: (1)Attention-based GCN auto-encoders are applied to better capture the relevance between nodes and neighbors for better node embeddings. (2)A contrastive fusion strategy is designed to better integrate the heterogeneous information. (3)MGCCN employs a self-supervised component that iteratively strengthens the node embedding and clustering. Extensive experiments on different types of real-world graph data demonstrate that our proposed method outperforms state-of-the-art techniques.

\section{Related Work} 
Early unsupervised graph representation learning methods are based on random walks \cite{perozzi2014deepwalk,grover2016node2vec} and encoder-decoder frameworks \cite{cao2016deep,tang2015line,wang2016structural}. Some approaches that take node attributes into account are also developed \cite{yang2015network,bojchevski2018bayesian,wang2016semantic}. Recently, a number of GCN-based methods are proposed. They targets to reconstruct the adjacency, e.g., GAE and VGAE \cite{kipf2016variational}, or the feature matrix \cite{wang2017mgae,park2019symmetric}. Adversarial regularizer is further introduced to enhance the robustness of GAE (ARGAE) \cite{pan2019adversarially}. Graph attention is also applied to capture the importance of the neighboring nodes to a target node (DAEGC) \cite{wang2019attributed}. Although these models have achieved impressive performance, they only focus on unilayer graph.

To cluster multilayer graph, some network embedding approaches have been proposed \cite{zhang2018scalable,liu2017principled}. O2MAC \cite{fan2020one2multi} makes the first attempt to use GCN. However, they are not flexible to handle multiview attributes. On the other hand, MAGCN \cite{cheng2020multi} is designed to cluster multiview atttibutes with a common graph. To simultaneously tackle multiview attributes and graphs, MvAGC is recently developed \cite{lin2021multi}. Nevertheless, it adopts a shallow model, which has a limited capacity. Motivated by these observations, we propose MGCCN, which combines GCN-based multilayer representation learning and self-supervised clustering through a contrastive funsion strategy.


\section{Methodology}
\subsection{Notation}
Let $\mathcal{G}$ be multilayer graph with $m$ layers and $N$ vertices denoted by
\begin{equation*}
    \mathcal{G} = \{G^s(V,E),X^s\}_{{1}\leq {s}\leq {m}},
\end{equation*}
where $G^s$ represents an undirected graph with vertex set $V$ and the set of edges $E$. The topology structure of graph $G^s$ is described by an adjacency matrix $\tilde{A}^s \in R^{N \times N}$, where element $\tilde{A}_{ij}=1$ if there is an edge between node $i$ and $j$. $X^s = [x_1,x_2,...,x_N]\in R^{d^s \times N}$ denotes the node attribute matrix with $d^s$ features.

\subsection{Self-Reconstruction}
As shown in Fig \ref{fig:model}, our framework is built upon GCN, which has demonstrated powerful graph representation learning ability.\\
\textbf{Encoder} We first need to learn a node representation $H_s\in{R^{N\times{F}}}$ for each layer. Specifically, we feed each adjacency matrix and attribute matrix to shared graph encoder to obtain graph embedding. Mathematically, for the $l$-th layer, GCN applies convolution on graph by spectral convolution function $f(H_s^{(l)},A^s;W^{(l)})$:

\begin{equation}
	\begin{aligned}
        H_s^{(l)} &= f(H_s^{(l-1)},A^s;W^{(l)}) \\
        &= \sigma(D_s^{-\frac{1}{2}}A^sD_s^{-\frac{1}{2}}H_s^{(l-1)}W^{(l)}),
	\end{aligned}
\end{equation}
where $A^s=\tilde{A}^s+I$, $I$ is an identify matrix, and $W^{(l)}$ is a matrix of trainable parameters. When $l=0$, $H_s^{(0)}=X^s$. $D_s$ is the degree matrix of $A^s$ with $i$-th diagonal element $\sum_{j}A_{ij}^s$, and $\sigma$ is an activation function. 

To better capture the relevance between nodes and neighbors, we employ attention mechanism \cite{salehi2019graph}. To be precise, the relevance of a neighboring node $j$ to node $i$ in the $s$-th graph can defined as:
\begin{equation}
	\begin{aligned}
        e_{ij}^{(l)} = sigmoid(c_1^{(l)}\sigma(W^{(l)}h_i^{(l-1)}) + c_2^{(l)}\sigma(W^{(l)}h_j^{(l-1)})),
	\end{aligned}
\end{equation}
where $h_i^{(l-1)}$ and $h_j^{(l-1)}$ denote $i$-th and $j$-th row of matrix $H_s^{(l-1)}$. $c_1^{l}$ and $c_2^{l}$ are both trainable parameters. $sigmoid$ refers to sigmoid function. Then, we normalize $e_{ij}$ to make relevance coefficients of node $i$’s neighbor set $\mathcal{N}_i$ comparable.
\begin{equation}
        a_{ij}^{(l)} = \frac{exp(e_{ij}^{(l)})}{\sum_{k\in{\mathcal{N}_i}}exp(e_{ik}^{(l)})}.
\end{equation}
Finally, we obtain the final embedding of node $i$:
\begin{equation}
   \label{equation:encoder}
        h_i^{(l)}=\sum_{j\in{\mathcal{N}_i}}a_{ij}^{(l)}\sigma(W^{(l)}h_j^{(l-1)}).
\end{equation}
\textbf{Decoder}. We use a symmetric decoder to reconstruct the representations of nodes by utilizing the embeddings of their neighbors. The decoder in $l-1$-th GCN layer can be computed as follows:
\begin{equation}
    \label{equation:decoder}
        \hat{h}_i^{(l-1)}=\sum_{j\in{\mathcal{N}_i}}\hat{a}_{ij}^{(l)}\sigma(\hat{W}^{(l)}\hat{h}_j^{(l)}),
\end{equation}
where $\hat{h}_i^{(l)}$ denotes the $i$-th row of matrix $\hat{H}_s^{(l)}$ in the decoder. $\hat{X}^s=\hat{H}_s^{(0)}$ is the output of last layer, which should be close to the input. Moreover, it is essential to preserve the graph structure information. Rather than reconstructing the adjacency matrix, which indicates that two unconnected nodes are dissimilarity, we require the representations of neighboring nodes similar \cite{salehi2019graph}. Particularly, the objective for $s$-th layer can be written as:
\begin{equation}
    L_{str}^s=-\sum^N_{i=1}\sum_{j\in{\mathcal{N}_i}}\log(\frac{1}{1+\exp(-{h}_i^{(l)}{h}_j^{(l)})})
\end{equation}

Finally, for all $m$ layers of inputs, the reconstruction loss can be formulated as follows:
\begin{equation}
    \label{equation:reconstruct}
    L_{re} = \sum_{s=1}^m\|X^s-\hat{X}^s\|_F^2 + \lambda_1 \sum_{s=1}^m L_{str}^s,
\end{equation}
where $\lambda_1>0$ is a trade-off parameter to balance attribute and graph reconstruction. By minimizing above reconstruction loss, the attention-based GCN is expected to capture the rich information in the graph and attribute.

\subsection{Contrastive Fusion}
As aforementioned, how to exploit heterogeneous information is challenging for multilayer graph clustering. In other words, we need to aggregate valuable information from each view and achieve an unified representation. For simplicity, we denote the output of encoder as $\{Z_s\}_{s=1}^{m}$. More recently, unsupervised representation learning with contrastive mechanism has reported promising performance. Specifically, contrastive learning aims to learn discriminative representations by contrasting positive and negative samples \cite{hassani2020contrastive,you2020graph}. In the literature, a number of methods have been developed to augment data and construct negative pairs, which are crucial to the performance of downstream tasks \cite{zhu2021graph,tian2020contrastive}. In this work, we employ a contrastive objective to fuse the embeddings without data augmentation. More precisely, we pull close the embeddings of the same node in different layers and push away the embedding of different nodes.

For any node $i$, its embedding produced in one view, $z_i$, is treated as the anchor. $z_i$ and the corresponding embeddings generated in other views, $z_i^{'}$, form the positive samples. $z_i$ and the embeddings of other nodes are regarded as negative pairs. The pairwise loss is defined below:
\begin{equation}
    \ell(z_i,z_i^{'}) = -\log \frac{e^{(sim(z_i,z_i^{'})/\tau)}}{\sum_{k=1}^Ne^{(sim(z_i,z_k^{'})/\tau)}+ \sum_{k=1}^Ne^{(sim(z_i,z_k)/\tau)}},
    \label{cont}
\end{equation}
where $\tau$ denotes the temperature parameter, $sim(z_i,z_k^{'})$ is the cosine similarity of inter-layer pairs, and $sim(z_i,z_k)$ represents the cosine similarity of intra-layer negative pairs. 
The overall objective to be minimized is the average of Eq.(\ref{cont}) over all positive pairs given by $ L(Z_s, Z_{s^{'}})= \frac{1}{2N}\sum_{i=1}^N[\ell(z_i,z_i^{'})+\ell(z_i^{'},z_i)]$. For all $m$ layers of inputs, the contrastive loss is formulated as follows:

\begin{equation}
    \label{equation:con}
    L_{con}= \sum_{s \neq s^{'}}^{m}L(Z_s, Z_{s^{'}})
\end{equation}
Afterwards, we combine latent representation of each view for clustering, $Z=\sum_{s=1}^m\beta_sZ_s$, which $\beta_s$ is a combination coefficient. 

\begin{table*}[htb]
    \caption{Statistics of the datasets.} 
    \label{tab:dataset}
    \centering
    \renewcommand\arraystretch{1.2}
\begin{tabular}{c|c|c|c|c|c|c}
\hline
\textbf{Datasets} &
  \textbf{Nodes} &
  \textbf{Relation Types} &
  \textbf{Edges} &
  \textbf{Feature Types} &
  \textbf{Attributes} &
  \textbf{Classes} \\ \hline \hline
  
\textbf{Cora} &
  2, 708 &
  Citation Network &
  5,429 &
  \begin{tabular}[c]{@{}c@{}} Bag of words of keywords \\ Cartesian transform\end{tabular} &
  \begin{tabular}[c]{@{}c@{}} 1, 433\\ 2, 708\end{tabular} &
  7 \\ \hline
\textbf{Citeseer} &
  3, 327 &
  Citation Network &
  4,732 &
 \begin{tabular}[c]{@{}c@{}} Bag of words of keywords \\ Cartesian transform\end{tabular} &
 \begin{tabular}[c]{@{}c@{}} 3, 703\\ 3, 327\end{tabular} &
  6 \\ \hline
\textbf{Wiki} &
  2, 405 &
  Webpage Network &
  17,981 &
   \begin{tabular}[c]{@{}c@{}} TF-IDF \\ Cartesian transform\end{tabular} &
  \begin{tabular}[c]{@{}c@{}} 4, 973\\ 2, 405\end{tabular} &
  17 \\ \hline
\textbf{ACM} &
  3, 025 &
  \begin{tabular}[c]{@{}c@{}}Paper-Subject-Paper(PSP)\\ Paper-Author-Paper (PAP)\end{tabular} &
  \begin{tabular}[c]{@{}c@{}}2,210,761\\ 29,281\end{tabular} &
    Bag of words of keywords &
  1, 830 &
  3 \\ \hline
\textbf{IMDB} &
  4, 780 &
  \begin{tabular}[c]{@{}c@{}}Movie-Actor-Movie(MAM)\\ Movie-Director-Movie(MDM)\end{tabular} &
  \begin{tabular}[c]{@{}c@{}}98,010\\ 21,018\end{tabular} &
  Bag of words of keywords &
  1, 232 &
  3 \\ \hline
\textbf{DBLP} &
  7, 907 &
  \begin{tabular}[c]{@{}c@{}}Paper-Author-paper(PAP)\\ Paper-Paper-Paper(PPP)\end{tabular} &
  \begin{tabular}[c]{@{}c@{}}144,783\\ 90,145\end{tabular} &
  Bag of words of keywords &
  2, 000 &
  4 \\ \hline
 \textbf{Amazon} &
  7, 621 &
  \begin{tabular}[c]{@{}c@{}}Item-AlsoView-Item (IVI)\\ Item-AlsoBought-Item (IBI) \\ Item-BoughtTogether-Item (IOI) \end{tabular} &
  \begin{tabular}[c]{@{}c@{}}266,237\\ 1,104,257 \\ 16,305  \end{tabular} &
  Bag of words of keywords &
  2, 000 &
  4 \\ \hline
\end{tabular}
\end{table*}

\subsection{Self-supervised Clustering}
Despite above procedures could output a high-quality embedding $Z$ of multilayer graph, there is no guarantee that it is optimal for clustering due to its unsupervised nature. Therefore, we propose to perform clustering in a unified framework, i.e., the embedding and clustering are updated iteratively. 
To bridge the embedding with clustering task, we use the Student’s t-distribution as a kernel to measure the similarity between each node and centroid \cite{xie2016unsupervised}:
\begin{equation}
    \label{equation:Q}
    q_{ij}         = \frac{(1+||z_i-\mu_j||^2)^{-1}}{\sum_{j^{\prime}=1}^k(1+||z_i-\mu_{j{\prime}}||^2)^{-1}},
\end{equation}
where $\{\mu_j\}^k_{j=1}$ is the $k$ initial cluster centroids. In fact, $q_{ij}$ is interpreted as the probability of assigning sample $i$ to cluster $j$, i.e., a soft assignment. Furthermore, an auxiliary target distribution $P$ is introduced to refine clusters as follows:
\begin{equation}
    \label{equation:P}
    p_{ij} = \frac{q_{ij}^2/f_j}{\sum_{j^{\prime}}q_{ij^{\prime}}^2/f_{j^{\prime}}},
\end{equation}
where $f_j=\sum_i{q_{ij}}$ is the soft cluster frequencies. Then, we can minimize the difference between soft clustering assignment and auxiliary target distribution as follows:
\begin{equation}
    \label{equation:clustering}
    L_{clu} = \|Q-P\|_F^2.
\end{equation}
To some extend, “highly confident” nodes serve as soft labels to supervise the clustering process.

\begin{algorithm}[tb]
\caption{MGCCN}
\label{alg:algorithm}
\textbf{Input}: The $m$ layers datasets with $N$ vertices, adjacency matrix $\tilde{A}^1,\cdots,\tilde{A}^m$, node attribute matrix  ${X}^1,\cdots,{X}^m$.\\
\textbf{Parameter}: Hyperparameter $\lambda_1$ ,  $\lambda_2$ , $\lambda_3$, parameter $\beta$, cluster number $g$\\
\textbf{Output}: $g$ partitions
\begin{algorithmic}[1] 
\STATE Let epoch $ = $ 0.
\WHILE{not converge or epoch $ \leq $ 1000}

\FOR{$l=1$ to $L$} 
\STATE {Compute $H_1^{(l)},\cdots,H_m^{(l)}$ according to (\ref{equation:encoder})}
\ENDFOR 

\STATE Capture consistent information from different layers according to (\ref{equation:con}) 

\STATE Aggregate each layer to achieve a unified representation $Z=\sum_{s=1}^m\beta_sZ_s$
\STATE Minimize $\|Q-P\|_F^2$ for self-supervised clustering
\FOR{$l=L$ to $1$} 
\STATE {Compute ${\hat{H}_1}^{(l)},\cdots,\hat{H}_m^{(l)}$ according to (\ref{equation:decoder}) } 
\ENDFOR 

\STATE Jointly train the overall network according to (\ref{finalmod})
\ENDWHILE
\STATE \textbf{return} label by (\ref{max})
\end{algorithmic}
\end{algorithm}

\begin{table}[!htb]
  \setlength\tabcolsep{3pt}
    \caption{Parameter values.}
    \label{tab:hyperparameters}
    \centering
    \renewcommand\arraystretch{1.1}
        \scalebox{0.95}{
\begin{tabular}{c|c|c|c|c}
\hline
\textbf{\textbf{Datasets}} &
  \textbf{Cora} &
  \textbf{Citeseer} & 
  \textbf{Wiki} & -
  \\ \hline
Network &
  [512 512] &
  [2000 512] &
  [512 512] & -
   \\ \hline
 \textbf{($\lambda_{1}, \lambda_{2}, \lambda_{3}$)} &
  (0.5, 10, 0.5) &
  (0.5, 10, 0.5) &
  (0.5, 10, 0.5) & -
  \\ \hline
  \textbf{($\beta_1$, $\beta_2$)} &
  (50, 1) &
  (10, 1) &
  (90, 1) & -

  \\ \hline
  \textbf{Datasets} &
  \textbf{ACM} & 
  \textbf{IMDB} & 
  \textbf{DBLP} &
  \textbf{Amazon}
    \\ \hline
 Network &
  [786 256] &
  [512 256] &
  [786 256] &
  [786 256]\\ \hline
  \textbf{($\lambda_{1}, \lambda_{2}, \lambda_{3}$)} &
  (0.05, 10, 0.5) &
  (0.05, 10, 0.5) &
  (0.05, 10, 0.5) & 
  (0.05, 50, 0.5) 
  \\ \hline
  \textbf{($\beta_1$, $\beta_2$)} &
  (1, 1)  &
  (1, 1) &
  (1, 1) &
  (1, 1, 1) 
  \\ \hline
  
\end{tabular}
}
\end{table}

\subsection{Overall Objective Function}
To sum up, MGCCN optimizes the following objective:
\begin{equation}
    L = \min{L_{re}} + \lambda_2 L_{con} + \lambda_3 L_{clu},
    \label{finalmod}
\end{equation}
where $\lambda_2$ and $\lambda_3$ are balance parameters. Therefore, MGCCN jointly trains GCN autoencoder, constrastive fusion and self-supervised clustering in an end-to-end fashion. When the network is well trained, the predicted cluster label can be inferred through last optimized $Q$, i.e.,
    \begin{equation}
    y_i = \mathop{\arg\max_{j}}q_{ij},
    \label{max}
\end{equation}
where $y_i$ is the cluster label of node $i$. The overall algorithm is outlined in Algorithm 1.

\begin{table*}[htb]
    \caption{Clustering results on multiview attribute datasets. ``bestX" means that the best results are chosen from different views. } 
    \label{tab:multi-attribute}
    \centering
    \renewcommand\arraystretch{1.2}
\begin{tabular}{c|c|c|c|c|c|c|c|c|c|c}
\hline
\multicolumn{1}{c|}{\multirow{2}{*}{\textbf{Method}}}  &
\multicolumn{1}{c|}{\multirow{2}{*}{\textbf{Input}}} &
\multicolumn{3}{c|}{Cora}  &
\multicolumn{3}{c|}{Citeseer} & 
\multicolumn{3}{c}{Wiki} \\ 
\cline{3-11}
&  & ACC    & NMI     &ARI     
& ACC    & NMI    & ARI 
& ACC    & NMI    & ARI 
   \\ \hline

K-means & $X^1$ & 0.500   & 0.317 & 0.239  & 0.544    & 0.312    & 0.285   &  0.417     &  0.440    &    0.151    \\ \hline
VGAE  \cite{kipf2016variational}  & $A\&\textit{bestX}$   
& 0.592    & 0.408   &  0.347   &  0.392   &    0.163  &  0.101    & 0.451   & 0.468            &  0.263   \\ \hline
GATE  \cite{salehi2019graph}  & $A\&\textit{bestX}$ & 0.658    & 0.527   &  0.451   &  0.616   & 0.401  &  0.381    & 0.482    & 0.343              &   0.188    \\ \hline
MGAE  \cite{wang2017mgae} & $A\&\textit{bestX}$  & 0.684    & 0.511   &  0.448   &  0.661   & 0.412  &  0.414        & 0.515        & \textbf{0.485}       & 0.349       \\ \hline
ARGAE \cite{pan2019adversarially} & $A\&\textit{bestX}$   & 0.640    & 0.449   & 0.352    & 0.573    & 0.350  &  0.341        & 0.381        & 0.345       & 0.112       \\ \hline
ARVGAE \cite{pan2019adversarially} & $A\&\textit{bestX}$   & 0.638    & 0.450   & 0.374    &  0.544   & 0.261  &  0.245        & 0.387        & 0.339       &  0.107       \\ \hline
DAEGC \cite{wang2019attributed} & $A\&\textit{bestX}$   &  0.704    & 0.528  & 0.496    &  0.672   & 0.397  &  0.410       & 0.521        & 0.432        &  0.337      \\ \hline
MAGCN \cite{cheng2020multi} & $A\&X^1\&X^2$   & 0.751    & 0.598   & 0.532    & 0.711    & \textbf{0.458}    
& 0.462  & 0.483   & 0.427  & 0.216  \\ \hline
MGCCN   & $A\&X^1\&X^2$   & \textbf{0.761}   & \textbf{0.602}  & \textbf{0.558}  & \textbf{0.715}    & 0.455    & \textbf{0.473}   & \textbf{0.539}   & 0.472  & \textbf{0.441}  \\ \hline
\end{tabular}
\end{table*}

\begin{table*}[!htbp]
    \centering
    \caption{Clustering results on multilayer graph datasets. ``X-avg" means that the results are based on the averaged node representation of each graph.}
    \label{tab:multi-graph}
    \renewcommand\arraystretch{1.2}
    \scalebox{0.78}{
\begin{tabular}{c|c|c|c|c|c|c|c|c|c|c|c|c|c|c|c|c}
\hline
\multirow{2}{*}{\textbf{Method}} & \multicolumn{4}{c|}{ACM}          & \multicolumn{4}{c|}{IMDB}         & \multicolumn{4}{c|}{DBLP}          &
\multicolumn{4}{c}{Amazon} \\ \cline{2-17} 
                        & ACC    & F1     & NMI    & ARI    & ACC    & F1  & NMI    & ARI    & ACC    & F1     & NMI    & ARI    & ACC    & F1     & NMI    & ARI \\ \hline
LINE  \cite{tang2015line}                 & 0.6479  & 0.6594 & 0.3941  & 0.3433 & 0.4268                              & 0.2870 & 0.0031 & -0.0090 &   -     &     -  &   -      & -  &   -     &     -  &   -      & -     \\ \hline
LINE-avg \cite{tang2015line}              & 0.6479 & 0.6594 & 0.3941  & 0.3432 & 0.4719                              & 0.2985 & 0.0063 & -0.0090 &   -     &   -     &    -    & - &   -     &     -  &   -      & -      \\ \hline
GAE \cite{kipf2016variational}            & 0.8216 & 0.8225 & 0.4914& 0.5444 & 0.4298                            & 0.4062 & 0.0402 & 0.0473  &    -    &   -     &  -      &  -  &   -     &     -  &   -      & -    \\ \hline
GAE-avg \cite{kipf2016variational}        & 0.6990 & 0.7025 & 0.4771  & 0.4378 & 0.4442                             & 0.4172 & 0.0413 & 0.0491 &    -    &    -    &   -    &   -  &   -     &     -  &   -      & -   \\ \hline

MNE  \cite{zhang2018scalable}             & 0.6370 & 0.6479 & 0.2999  & 0.2486 & 0.3958                              & 0.3316 & 0.0017 & 0.0008  &    -   &   -     &  -       &   -  &   -     &     -  &   -      & -   \\ \hline
PMNE(n)  \cite{liu2017principled}               & 0.6936 & 0.6955 & 0.4648  & 0.4302 & 0.4958                              & 0.3906 & 0.0359 & 0.0366  &    -   &   -     &  -       & -   &   -     &     -  &   -      & -    \\ \hline
PMNE(r)  \cite{liu2017principled}               & 0.6492 & 0.6618 & 0.4063  & 0.3453 & 0.4697                              & 0.3183 & 0.0014 & 0.0115  &  -     &  -      &    -     & -  &   -     &     -  &   -      & -     \\ \hline
PMNE(c)  \cite{liu2017principled}               & 0.6998 & 0.7003 & 0.4775  & 0.4431 & 0.4719                              & 0.3882 & 0.0285 & 0.0284  &  -     &  -      &  -       &  -  &   -     &     -  &   -      & -    \\ \hline
RMSC   \cite{xia2014robust}                 & 0.6315 & 0.5746 & 0.3973  & 0.3312 & 0.2702                              & 0.3775 & 0.0054 & 0.0018  &     -  &   -     &    -     &  -   &   -     &     -  &   -      & -   \\ \hline
PwMC   \cite{nie2017self}                 & 0.4162 & 0.3783 & 0.0332  & 0.0395 & 0.2453                              & 0.3164 & 0.0023 & 0.0017  &    -   &     -   &  -       & -   &   -     &     -  &   -      & -    \\ \hline
SwMC   \cite{nie2017self}                 & 0.3831 & 0.4709 & 0.0838  & 0.0187 & 0.2671                              & 0.3714 & 0.0056 & 0.0004  &  -     &    -    &   -      &   -    &  -     &    -    &   -      &   -    \\ \hline
O2MAC  \cite{fan2020one2multi}           & 0.9042 & \textbf{0.9053} & 0.6923 & 0.7394 & 0.4502 & 0.4159 & 0.0421 & 0.0564 & 0.7267 & 0.7320 & 0.4066 & 0.4036 & 0.4428 & 0.4424 & 0.1344 & 0.0898 \\ \hline
MvAGC  \cite{lin2021graph}                 & 0.8975 & 0.8986 & 0.6735 & 0.7212  & \textbf{0.5633} & 0.3783 & 0.0371 & 0.0940 & {0.7221}  & 0.7332 &{0.4191}
&  0.4049 & 0.5188 & \textbf{0.5072} & \textbf{0.2322} & 0.1141  \\ \hline
HAN  \cite{wang2019heterogeneous}                  & 0.8823 & 0.8844 & 0.5881 & 0.5933  & 
0.5547 & 0.4152 & \textbf{0.0986} & 0.0856 & {0.7615}  & 0.6309 &{0.4866}
&  0.4635 & 0.4355 & 0.4246 & 0.1120 &0.0362 \\ \hline
MGCCN                 & \textbf{0.9167}  & 0.8472        & \textbf{0.7095}        & \textbf{0.7688}       & 0.5490      & \textbf{0.4740}        & 0.0567       & \textbf{0.1071}     & \textbf{0.8301} & \textbf{0.7336} & \textbf{0.6156} & \textbf{0.5976} & \textbf{0.5309} & 0.4572  & 0.1931 & \textbf{0.1860}\\ \hline

\end{tabular}}
\end{table*}

\section{Experiment}

\subsection{Dataset}
To assess the generic and effective of our model, we conduct experiments on multiple real-world datasets from different domains. We choose two citation network datasets widely used in related works: Cora \cite{mccallum2000automating} and Citseer \cite{giles1998citeseer}. Their nodes denote publications and the edges represent citations. Wiki \cite{yang2015network} is a webpage network and the nodes denote webpages; two nodes are connected if they link each other. These three data only have a single-layer. Following \cite{cheng2020multi}, we construct the second attribute with Cartesian product. Consequently, two attribute views shows the same graph. Besides, we also conduct experiments on another three popular datasets \cite{fan2020one2multi}: ACM, DBLP, and IMDB. There are two-layers that share one attribute. Specifically, the ACM dataset is a paper network with two types of paper relation: co-paper (two papers are written by the same author) and co-subject (two papers contain the same subjects). The DBLP dataset is a multiplex network with two relations: paper-author-paper, paper-paper-paper. The IMDB is a two-layer movie graph dataset, including movie-actor-movie and movie-director-movie relations. In addition, we also test a three-layer network: Amazon, which has four categories and includes three relations (Also View, Also Bought and Bought Together). Here, all graphs share the same attributes.
The detailed statistics of the datasets are shown in Table \ref{tab:dataset}.

\subsection{Experimental Settings}

In our network architecture, different views of graph share the same autoencoder parameters. We use a two-layer GCN encoder with latent dimension searched in range of $[32, 64, 128, 256, 512, 786, 1024, 2048]$. The learning rate is set to 0.003 for autoencoder-based models optimized with Adam algorithm. We tune $\lambda_{1}$, $\lambda_{2}$, $\lambda_{3}$ in the range of $[0.01, 0.5, 1, 10, 100]$. To train our model, we set the maximum iteration number to 1000, and stop our program when the loss and accuracy are close to stable. Non-linear activation function sigmoid is applied in the network. The used
parameters for the reported results are shown in Table \ref{tab:hyperparameters}. The most frequently used evaluation metrics are adopted to evaluate the performance: accuracy (ACC), normalized mutual information (NMI), F-score (F1), and average rand index (ARI). For a fair comparison, we copy part of the results from \cite{fan2020one2multi,cheng2020multi}.


\subsection{Compared Methods}
For a comprehensive evaluation, a number of methods from different categories are compared.
\begin{itemize}
    \item \textbf{K-means}: a classical clustering algorithm for feature data. 
    \item \textbf{GAE-based techniques}: graph autoencoder and its variational variant (VGAE) \cite{kipf2016variational}, marginalized graph autoencoder (MGAE) \cite{wang2017mgae}, adversarial regularized graph autoencoder (ARGAE) and its variational variant (ARVGAE) \cite{pan2019adversarially}, deep attentional embedding graph clustering (DAEGC) \cite{wang2019attributed}, and graph attention autoencoder (GATE) \cite{salehi2019graph}. They can only process single-layer graph, thus we run them on each attribute and report the best performance.
    \item \textbf{LINE \cite{tang2015line}}: a classical network embedding method.
    \item \textbf{MNE \cite{zhang2018scalable}}: it learns a unified embedding for multiplex network.
    \item \textbf{PMNE \cite{liu2017principled}}: it processes multilayer network through network aggregation (PMNE(n)), results aggregation (PMNE(r)), layer co-analysis (PMNE(c)). 
    \item \textbf{RMSC \cite{xia2014robust}}: a multi-view spectral clustering method based on Markov chain.  
    \item \textbf{PwMC and SwMC \cite{nie2017self}}: PwMC is a parameter-weighted multiview graph clustering method and SwMC corresponds to a self-weighted variant.
    \item \textbf{MAGCN \cite{cheng2020multi}}: a multiview attributed graph clustering method based on GCN.
    \item \textbf{O2MAC \cite{fan2020one2multi}}: it clusters multilayer graph by selecting an informative graph.
    \item \textbf{MvAGC \cite{lin2021graph}}: it proposes a shallow model to cluster data that have both multiview attributes and multiple graphs. It has achieved promising performance by using a graph filtering technique to smooth the signal and exploring high-order relations.
    \item \textbf{HAN \cite{wang2019heterogeneous}}: it is a representative GNN method for heterogeneous graph that contains different types of nodes and links. It  learns node embeddings by aggregating features from meta-path based neighbors in a hierarchical attention.
\end{itemize}

\begin{figure*}[htb]
	\begin{minipage}{0.15\linewidth}
		\vspace{3pt}
		\centerline{\includegraphics[width=\textwidth]{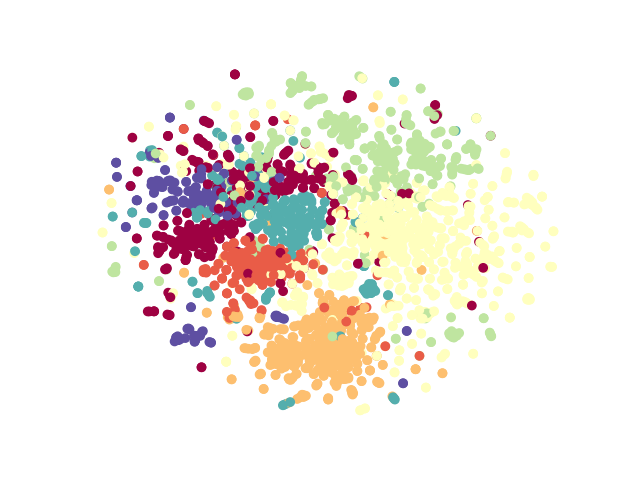}}
		\caption*{Cora: epoch 0}
	\end{minipage}
	\hfill
	\begin{minipage}{0.15\linewidth}
		\vspace{3pt}
		\centerline{\includegraphics[width=\textwidth]{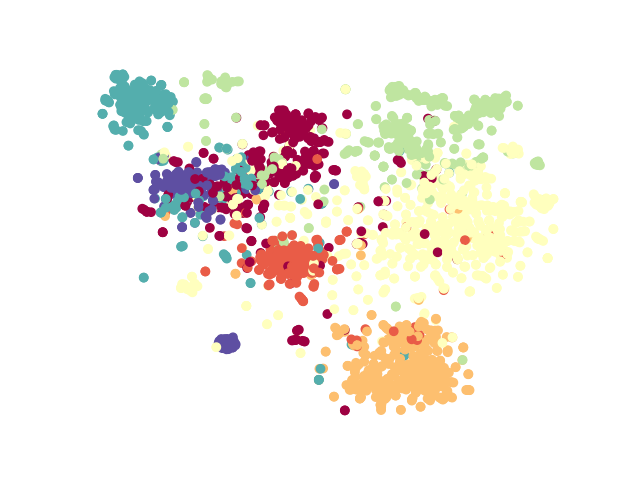}}
				\caption*{Cora: epoch 400}
	\end{minipage}
	\hfill
	\begin{minipage}{0.15\linewidth}
		\vspace{3pt}
		\centerline{\includegraphics[width=\textwidth]{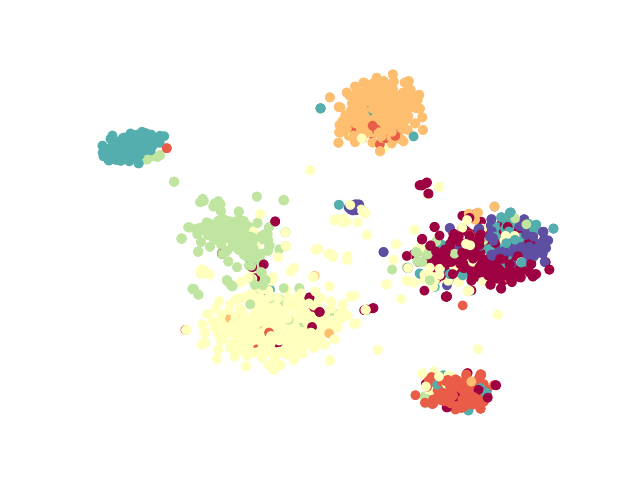}}
				\caption*{Cora: epoch 700}
	\end{minipage}
	\hfill
	\begin{minipage}{0.15\linewidth}
		\vspace{3pt}
		\centerline{\includegraphics[width=\textwidth]{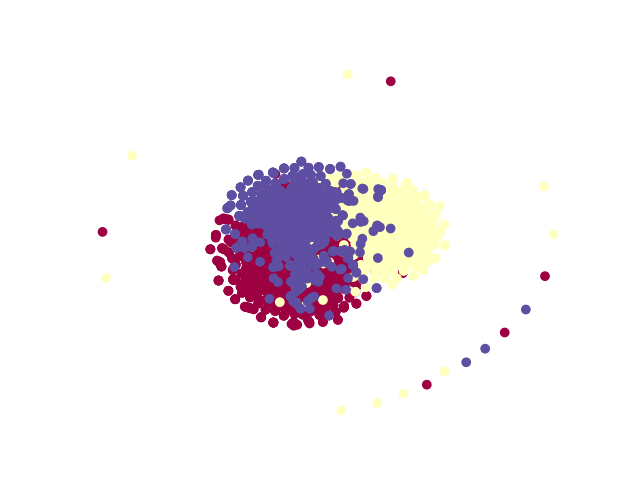}}
				\caption*{ACM: epoch 0}
	\end{minipage}
	\hfill
	\begin{minipage}{0.15\linewidth}
		\vspace{3pt}
		\centerline{\includegraphics[width=\textwidth]{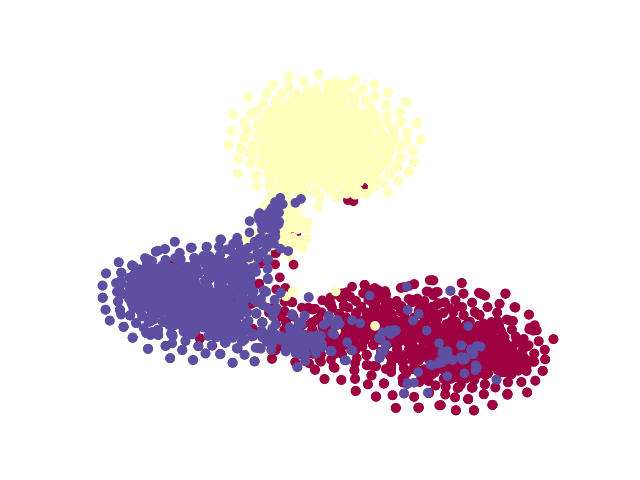}}
		\caption*{ACM: epoch 60}
	\end{minipage}
	\hfill
	\begin{minipage}{0.15\linewidth}
		\vspace{3pt}
		\centerline{\includegraphics[width=\textwidth]{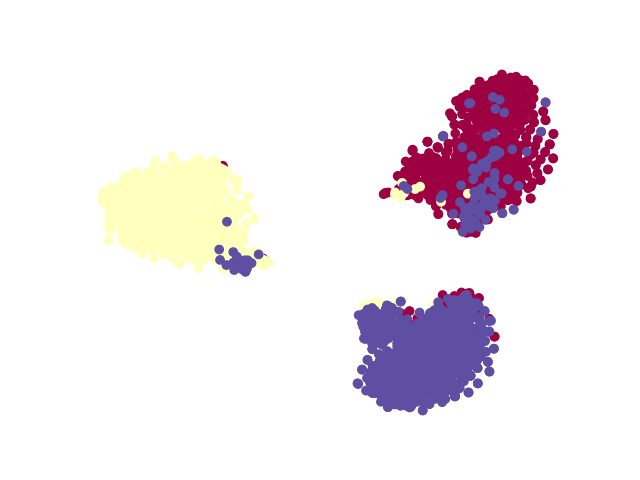}}
				\caption*{ACM: epoch 100}
	\end{minipage}
	\hfill
	
	\caption{The 2D visualization of embeddings on two datasets using t-SNE. The clusters are denoted by different colors.}
	\label{figure:TSNE}
\end{figure*}

\begin{figure*}[!htb]
\includegraphics[width=0.24\linewidth]{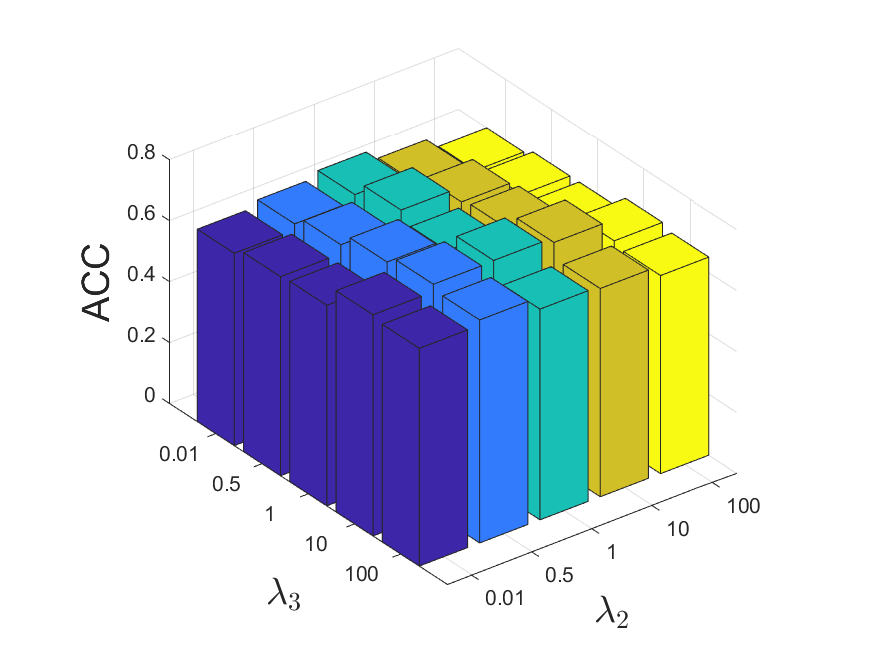}
\includegraphics[width=0.24\linewidth]{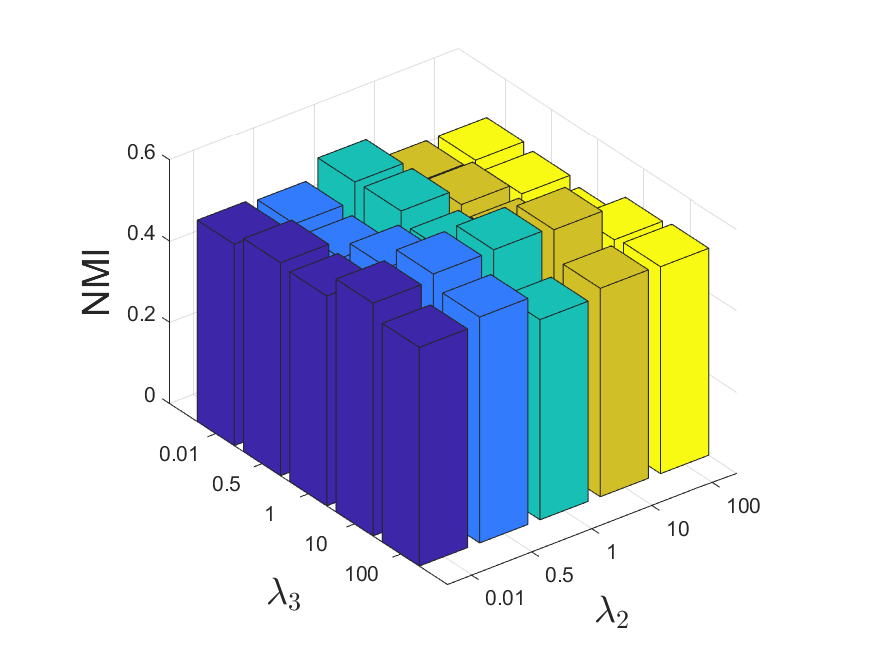}
\includegraphics[width=0.24\linewidth]{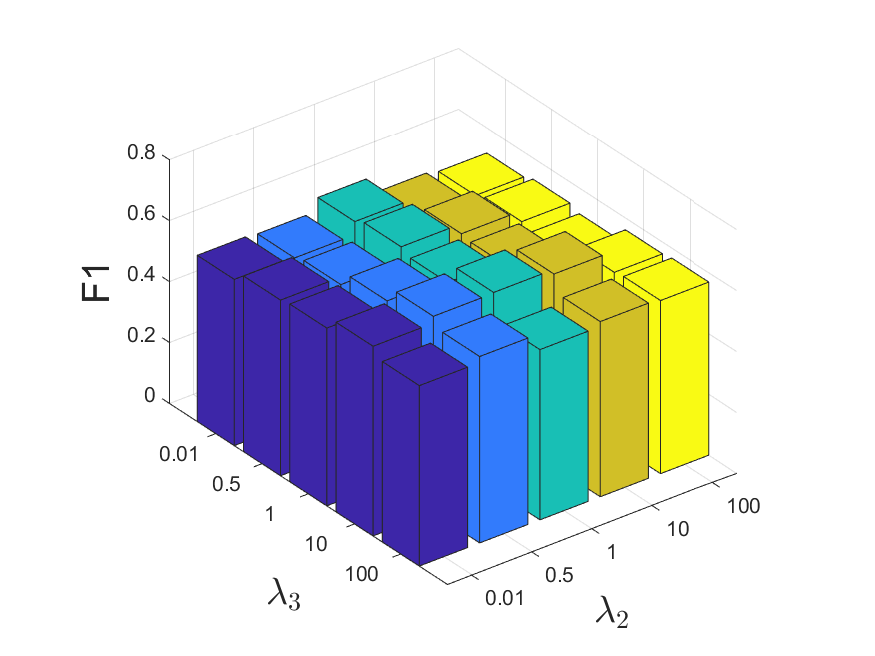}
\includegraphics[width=0.24\linewidth]{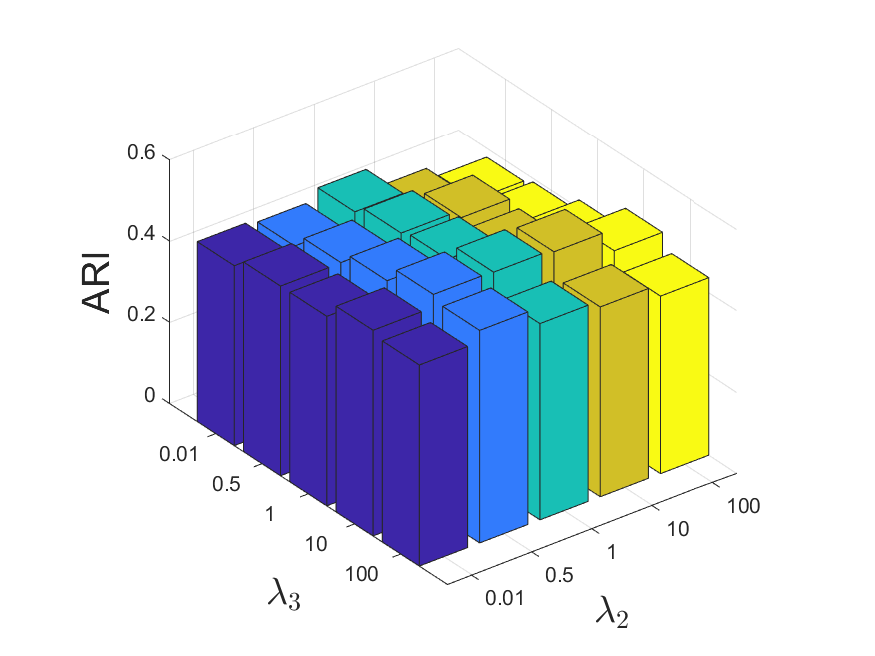}\\
\includegraphics[width=0.24\linewidth]{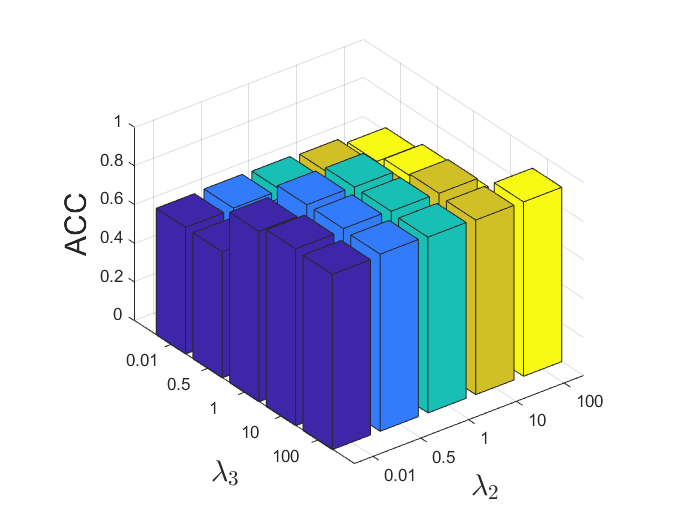}
\includegraphics[width=0.24\linewidth]{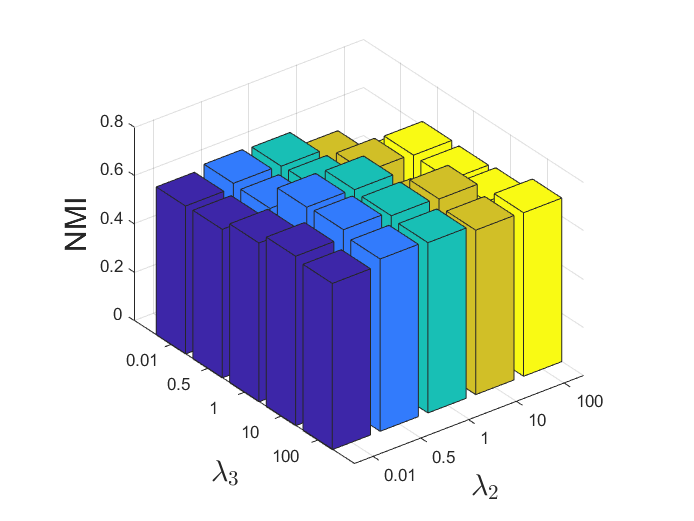}
\includegraphics[width=0.24\linewidth]{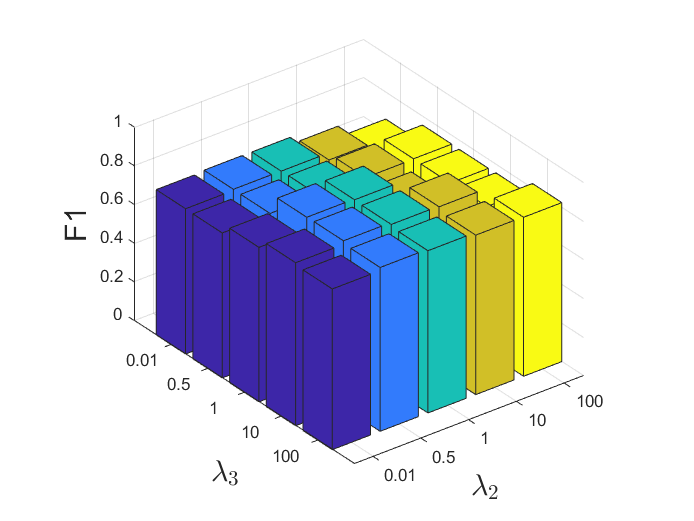}
\includegraphics[width=0.24\linewidth]{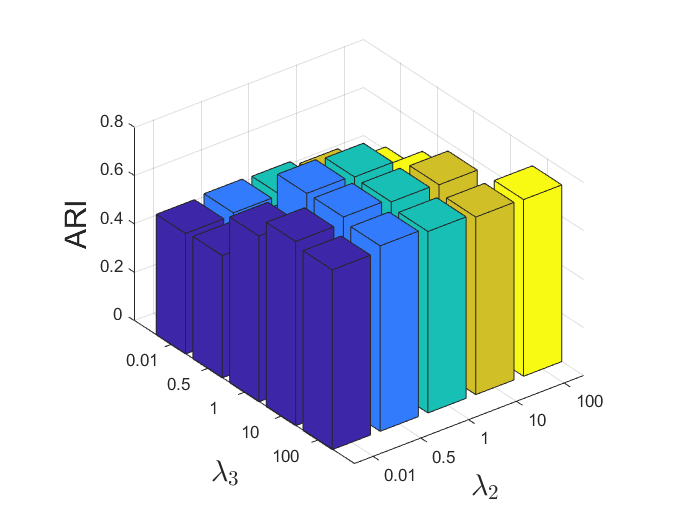}
\caption{The influence of $\lambda_2$ and $\lambda_3$ on the performance of Cora (1st row) and ACM (2nd row).}
	\label{parameter-analysis}
\end{figure*}

\subsection{Experimental Results}
The results on the seven datasets are given in Tables \ref{tab:multi-attribute} and \ref{tab:multi-graph}. For multilayer graph data, the results of MGCCN outperform the other approaches in most cases and the improvement is more than 10\% on some metrics. For multiview attribute data, the improvement is around 1\%. 
Specifically, we can draw the following conclusions. 
\begin{itemize}
\item Our proposed method demonstrates advantages over MAGCN and O2MAC, which are purposely designed for multiview attributes and multilayer graph clustering. Different from them, our method is a generic framework to handle both scenarios. Moreover, we introduce contrastive learning mechanism to boost the embeddings. 
\item With respect to recent method MvAGC, MGCCN also shows superior performance. This is partially attributed to the adoption of end-to-end deep neural network framework in MGCCN. Besides, contrastive idea is also crucial for representation learning. 
    \item Multiview methods often surpass single-view methods, which meets our expectation since they leverage information from multiple sources. For example, MGCCN and MAGCN show a clear edge over GAE-based methods in Table \ref{tab:multi-attribute}. 
    \item Techniques exploiting both attribute and graph structure information always report better performance than that of methods rely on one type of input. This validates that attribute and topology information contribute to clustering from different perspectives.
    \item Compared to heterogeneous graph neural network method HAN, our method also shows advantages in most cases. This benefit is mainly from our introduction of self-supervision mechanism. 
\end{itemize}
To intuitively show the embedding quality, we implement t-SNE on the learned representation $Z$ for visualization in Fig \ref{figure:TSNE}. We can see that MGCCN can well cluster the nodes after some iterations. In particular, the clusters have fewer overlapping areas as the training epoch increases.
\begin{table}[!htbp]
 \centering
    \setlength{\abovecaptionskip}{0pt}%
    \setlength{\belowcaptionskip}{10pt}
    \caption{Ablation study of the model.} 
    \label{tab:wo}
    \renewcommand\arraystretch{1.2}
    \scalebox{0.95}{
\begin{tabular}{c|c|l|l|l}
\hline
\multicolumn{1}{l|}{\textbf{Datasets}} &
  \multicolumn{1}{l|}{Metrcis} &
  MGCCN &
  \multicolumn{1}{c|}{\begin{tabular}[c]{@{}c@{}}MGCCN\\ (w/o $L_{con}$)\end{tabular}} &
  \multicolumn{1}{c}{\begin{tabular}[c]{@{}c@{}}MGCCN\\ (w/o $L_{clu}$)\end{tabular}} \\ \hline
\multirow{4}{*}{Cora}     & ACC & \multicolumn{1}{c|}{0.761} & \multicolumn{1}{c|}{0.745}  & \multicolumn{1}{c}{0.710}  \\                                  \cline{2-5} 
 & NMI & \multicolumn{1}{c|}{0.602} & \multicolumn{1}{c|}{0.581}  & \multicolumn{1}{c}{0.548} \\ \cline{2-5} 
                          & ARI & \multicolumn{1}{c|}{0.558} & \multicolumn{1}{c|}{0.539} & \multicolumn{1}{c}{0.497}  \\ \hline
\multirow{4}{*}{Citeseer} & ACC & \multicolumn{1}{c|}{0.715} & \multicolumn{1}{c|}{0.696} &   \multicolumn{1}{c}{0.304} \\ \cline{2-5} 
                          & NMI & \multicolumn{1}{c|}{0.455} & \multicolumn{1}{c|}{0.432} &  \multicolumn{1}{c}{0.143} \\ \cline{2-5} 
                          & ARI & \multicolumn{1}{c|}{0.473} & \multicolumn{1}{c|}{0.447} &  \multicolumn{1}{c}{0.057}  \\ \hline \hline
\multirow{4}{*}{ACM}      & ACC &  \multicolumn{1}{c|}{0.9167}  & \multicolumn{1}{c|}{0.9107}  & \multicolumn{1}{c}{0.6525} \\ \cline{2-5} 
                          & F1  & \multicolumn{1}{c|}{0.8472}   & \multicolumn{1}{c|}{0.8376} &  \multicolumn{1}{c}{0.7526} \\ \cline{2-5} 
                          & NMI &  \multicolumn{1}{c|}{0.7095}  & \multicolumn{1}{c|}{ 0.6958} & \multicolumn{1}{c}{0.6151} \\ \cline{2-5} 

                          & ARI & \multicolumn{1}{c|}{0.7688}   & \multicolumn{1}{c|}{ 0.7539} & \multicolumn{1}{c}{0.4959} \\ \hline
\multirow{4}{*}{IMDB}     & ACC & \multicolumn{1}{c|}{0.5490}   & \multicolumn{1}{c|}{0.5244} & \multicolumn{1}{c}{0.4685} \\ \cline{2-5} 
                          & F1  & \multicolumn{1}{c|}{0.4740}   & \multicolumn{1}{c|}{0.4602} &  \multicolumn{1}{c}{0.4415}\\ \cline{2-5} 
                          & NMI & \multicolumn{1}{c|}{0.0567}   & \multicolumn{1}{c|}{0.0443} &  \multicolumn{1}{c}{0.0114}\\ \cline{2-5} 
                          & ARI & \multicolumn{1}{c|}{0.1071}   & \multicolumn{1}{c|}{0.1094} &  \multicolumn{1}{c}{0.0121}\\ \hline
\end{tabular}}
\end{table}

\begin{figure}[h]
    \centering
    \includegraphics[width=.4\textwidth]{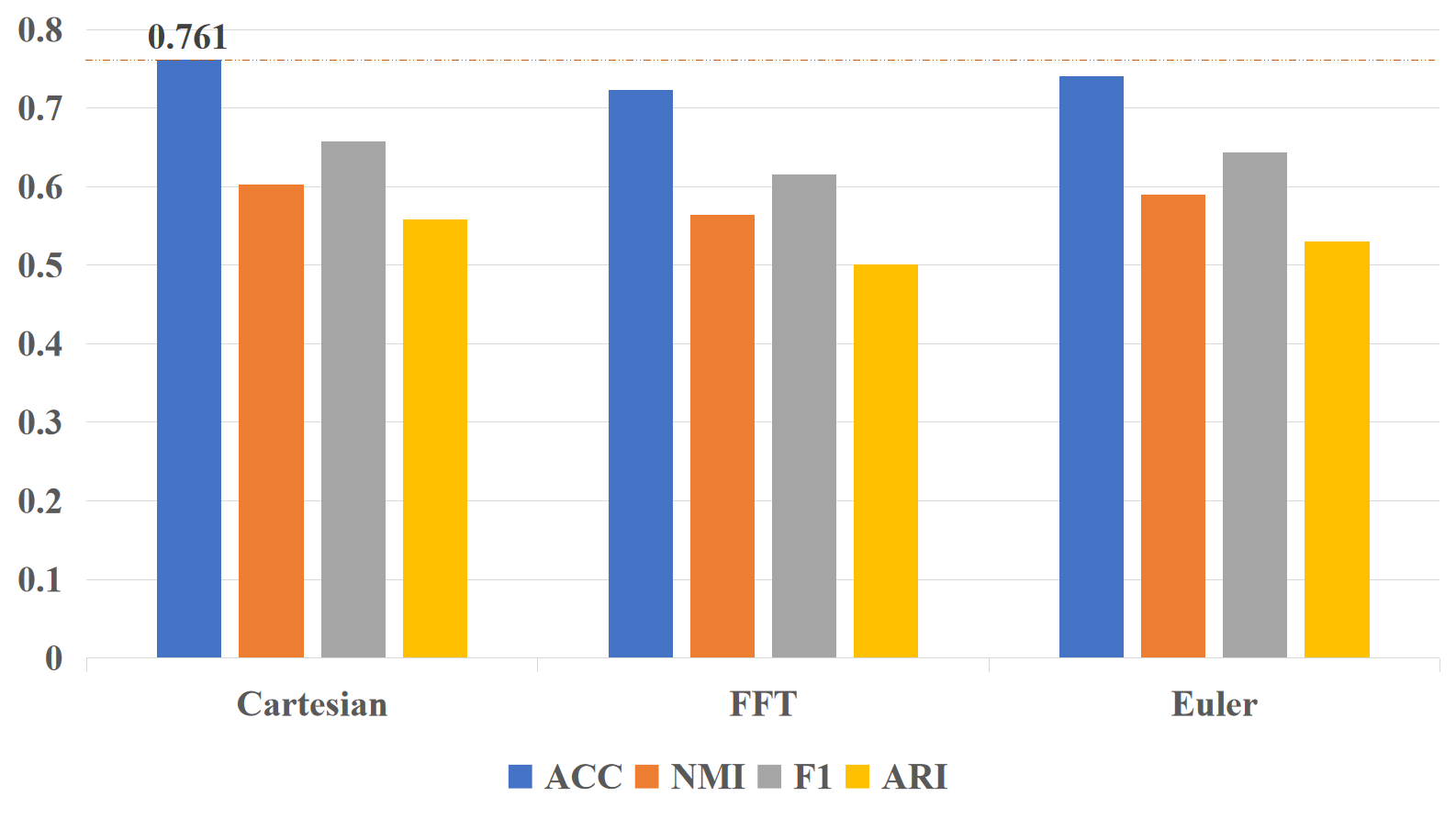}
    \caption{The performance of Cora with attribute constructed by different methods.}
    \label{fig:ablation—multi-attribute-la}
\end{figure}

\subsection{Ablation Study}
To validate the effectiveness of different components in our model (\ref{finalmod}), we perform ablation study on each module. Particularly, we test this by removing $L_{con}$ and $L_{clu}$ respectively.
As shown in Table \ref{tab:wo}, it is obvious that both contrastive fusion and self-supervised clustering are important to the performance. 
In addition, we assess the influence of transformation in multiview attribute case on Cora. To this end, we further adopt Fast Fourier transform (FFT) and Euler transform (Euler) to form the second attribute $X^{2}$. From Fig \ref{fig:ablation—multi-attribute-la}, we can see that MGCCN achieves the best clustering performance with Cartesian product.

Furthermore, we evaluate the effect the multiview attributes or graph by removing some information. According to Table \ref{tab:single-layer}, deleting the 2nd attribute in Cora or the 2nd graph in ACM degrades the performance. This verifies the significance of considering all information from different sources.

\begin{table}[htb]
    \setlength{\abovecaptionskip}{0pt}%
    \setlength{\belowcaptionskip}{10pt}
    \caption{The test of attribute and graph.} 
    \label{tab:single-layer}
    \centering
    \renewcommand\arraystretch{1.2}
\begin{tabular}{c|c|c|c|c|c}
\hline
\multicolumn{2}{c|}{\textbf{Datasets}} &
  ACC &
  NMI & 
  F1 &
  ARI 

  \\ \hline
     \multirow{2}{*}{ACM} &$A^1,X$ & 0.8997 &0.6723 & 0.8197 & 0.7273\\ \cline{2-6}
     &$A^1,A^2,X$ & 0.9167 & 0.7095&0.8472 & 0.7688
    \\ \hline
   \multirow{2}{*}{Cora} &$A,X^1$& 0.6890 & 0.5586 &0.6010 &0.4841\\ \cline{2-6}
   &$A,X^1,X^2$& 0.7610 & 0.6022 & 0.6585 &0.5589
    \\ \hline
\end{tabular}
\end{table}

\subsection{Parameter Analysis}
There are three regularization parameters in our model: $\lambda_{1}$, $\lambda_{2}$, and $\lambda_{3}$. Taking Cora and ACM for examples, we show the parameter sensitivity in Fig \ref{parameter-analysis}. To this end, we fix $\lambda_{1}$ as in Table \ref{tab:hyperparameters}. Afterwards, we tune the values of $\lambda_{2}$ and $\lambda_{3}$. It can be observed that our method produces reasonable results in a wide range of $\lambda_{2}$ and $\lambda_{3}$. In particular, small $\lambda_{3}$ and big $\lambda_{2}$ both degrade the performance. On the other hand, we can fix one of them and tune the other without sacrificing the performance. 

\section{Conclusion}
In this paper, we propose a model for multilayer graph clustering task. We design a GCN-based architecture to flexibly process multilayer graph and multiview attributes. We further propose a contrastive fusion module to capture the consistency information among different layers. A self-supervised clustering module is also applied to achieve high-quality result. Comprehensive experimental results indicate that our proposed method outperforms other state-of-the-art techniques. 

\bibliographystyle{IEEEtran}
\bibliography{bare_jrnl}



%








\end{document}